\documentclass[sigconf, nonacm=true]{acmart}

\usepackage[english]{babel}
\usepackage{blindtext}

\RequirePackage{booktabs}

\renewcommand\footnotetextcopyrightpermission[1]{} % removes footnote with conference info
\setcopyright{none}

\begin{document}
\title{A Pipeline for DNS-Based Software Fingerprinting}

\author{Sebastian Schäfer}
\affiliation{
 \institution{RWTH Aachen University}
}
\email{schaefer@itsec.rwth-aachen.de}

\author{Ulrike Meyer}
\affiliation{
 \institution{RWTH Aachen University}
}
\email{meyer@itsec.rwth-aachen.de}

% The default list of authors is too long for headers}
\renewcommand{\shortauthors}{Schäfer et al.}

\begin{abstract}
%The total number of internet-connected devices has been tremendously increasing over the last decade. Common to all of these devices is the use of the 30 years old DNS protocol that mainly translates domain names to IP addresses. As DNS is used prior to most internet connections and is often unencrypted, passively monitoring DNS traffic reveals much information on the devices used in a network as well as the software installed on them.
 
In this paper, we present the modular design and implementation of DONUT, a novel tool for identifying software running on a device. Our tool uses a rule-based approach to detect software-specific DNS fingerprints (stored in an easily extendable database) in passively monitored DNS traffic. We automated the rule extraction process for DONUT with the help of ATLAS, a novel tool we developed for labeling network traffic by the software that created it. We demonstrate the functionality of our pipeline by generating rules for a number of applications, evaluate the performance as well as scalability of the analysis, and confirm the functional correctness of DONUT using an artificial data set for which the ground-truth is known. In addition, we evaluate DONUT's analysis results on a large real-world data set with unknown ground truth. 
%\todo[inline]{TODO: first version. Revisit when eval chapter is done}
 
\end{abstract}

\maketitle

\section{Introduction}

As networks today become increasingly complex it becomes more and more important for network administrators, security consultants, and pentesters to quickly and unintrusively  obtain an overview on the devices running in a specific network in an automated fashion. Besides the general network topology, the information of interest includes identifying the operating systems devices run on, as well as the applications installed on them. An administrator in an open network such as a university network can typically not install administrative tools on every device used in the network. Nevertheless, he may, e.g., be interested in statistics related to the general use of operating systems and applications used in his network. Also, he may be interested in checking certain policies on which applications are prohibited to be installed on self-administrated devices, e.g., cloud storage applications due to privacy concerns or crypto mining software. Security consultants and pentesters may also be interested in this information in order to determine potential targets within a network. 

There are a variety of network scanning and mapping tools such as \cite{nmap, prtg, advip} that enable the generation of such an overview. These tools are either based on actively scanning the network or based on deep packet inspection and are thus rather complex and application specific. The same holds for other active approaches for fingerprinting operating systems based on ICMP \cite{arkin2002remote} or TCP SYN packets \cite{greenwald2007toward}. 
Passive approaches for desktop computers so far focus on fingerprinting operating systems (rather than applications) and are based on flows \cite{mossel2010passive,jirsik2014identifying}, TCP/IP header information \cite{lippmann2003passive}, TLS handshakes \cite{husak2016https,garn2019browser}, or DNS traffic \cite{matsunaka2013passive,chang2015study}. Only on  the mobile side fingerprinting of apps in general \cite{taylor2016appscanner,miskovic2015appprint,dai2013networkprofiler,van2020flowprint} has received some attention. Fingerprinting desktop applications based on passive monitoring has not yet been explored beyond fingerprinting DNS server software \cite{chitpranee2013towards}, browsers \cite{garn2019browser}, broader process families \cite{anderson2019tls}, and malware \cite{anderson2020accurate}. 

In this paper we present the design and implementation of a pipeline for rule-based application fingerprinting, including tools for automated network traffic labeling, rule-extraction, and analysis of passively monitored DNS traffic. The Domain Oriented Network Unmasking Tool (DONUT) covers the analysis part of the pipeline and identifies applications running on devices based on rules that represent application-specific patterns of domains queried by the devices. The Rule-Extractor automatically extracts rules for DONUT from application-labeled DNS traffic and obtains the labeled traffic from our novel Automated Traffic Labeling Software (ATLAS). Note that ATLAS is of interest independent of the context of the pipeline presented in this paper as it allows for labeling any network traffic on a Windows host with the application that initiated the traffic on a per-packet basis. 
We demonstrate the rule extraction process on a variety of commonly used applications and evaluate the performance and functional correctness of DONUT's analyzing capabilities on a self-generated dataset with known ground truth. We illustrate DONUT's real-world applicability by discussing its analysis results on a large-scale real world dataset from an open university network for which the ground truth is unknown.

\section{Related Work}
\label{chap:related_work}

%DNS zone enumeration \citeauthor{rose2009information} \cite{rose2009information}
%goal: obtain entire database of a zone

%privacy preserving dns \citeauthor{lu2010towards} \cite{lu2009ppdns,lu2010towards}
%\cite{ietfdnsprivacy,federrath2011privacy,hu2016specification,wachs2014censorship}
%approaches to mitigate the privacy leakage of DNS

%dns server fingerprinting \citeauthor{chitpranee2013towards}
%os fingerprinting \citeauthor{matsunaka2013passive} \cite{matsunaka2013passive}
%os fingerprinting based on dns log analysis \citeauthor{chang2015study} \citep{chang2015study}

%app fingerprinting \citeauthor{taylor2016appscanner} \citeauthor{miskovic2015appprint} \citeauthor{dai2013networkprofiler}

%nmap as an active approach
%ettercap?
Fingerprinting operating systems or application software has been studied using a multitude of different approaches. These approaches can broadly be separated into approaches based on actively scanning the network and those that are based on passive monitoring. In addition, they differ greatly in what they fingerprint (operating systems, or specific or general applications) based on which type of traffic (e.g.\ HTTP or SSL/TLS). In the following, we first discuss prior approaches for passive fingerprinting of operating systems and then carry on with the passive approaches for fingerprinting specific (mobile) applications. We then briefly discuss active fingerprinting approaches and place our work in the more general context of DNS privacy.  

%\medskip
%\noindent
%\textbf{Passive OS fingerprinting:}
Most closely related to DONUT is the passive approach to fingerprinting OSs based on DNS traffic presented in \cite{matsunaka2013passive}. It is based on characteristic queries and timing patterns. Another approach based on characteristic domain names in DNS queries is \cite{chang2015study}, where machine learning techniques are used to fingerprint OSs. Both of these approaches are similar to our approach in that they focus on passively monitored traffic. However, as opposed to DONUT these approaches focus on operating systems alone, while DONUT also fingerprints applications. 
%In \cite{wills2003inferring}, the authors use the idea of characteristic domains of internet applications to collect information about their popularity by periodically querying DNS servers to determine whether these domains are cached.

%The other passive approaches for OS fingerprinting, focus on monitoring protocols other than DNS and are thus less well suited to be extended to fingerprinting applications in general. For example, in \cite{husak2016https} user agent fingerprinting based on SSL/TLS handshakes is used to determine the OS of a device. Specifically they determine the cipher suite list as the most reliable parameter for OS detection. Extracting features for OS classification from passive analysis of TCP/IP header information to train different machine learning classifiers like binary trees is suggested in \cite{lippmann2003passive}. \cite{zhang2009remote} uses Support Vector Machines for OS detection and discovery of unknown fingerprints. In \cite{aksoy2017operating} a method for automatic detection of packet features for OS fingerprinting is presented. The approach is adaptive and allows to perform classification on any type of packet. Other authors use network flow analysis for OS fingerprinting. In \cite{mossel2010passive} the general feasibility of identifying an operating systems by monitoring network flows is shown. The authors of \cite{jirsik2014identifying} present a framework for passive OS detection based on flows and describe its deployment in an university network.

The other passive approaches for OS fingerprinting focus on monitoring protocols other than DNS and are thus less well suited to be extended to fingerprinting applications in general. For example, in \cite{husak2016https} user agent fingerprinting based on SSL/TLS handshakes is used to determine the OS of a device. Extracting features for OS classification from passive analysis of TCP/IP header information was suggested in \cite{lippmann2003passive}. \cite{zhang2009remote} uses Support Vector Machines for OS detection and discovery of unknown fingerprints. In \cite{aksoy2017operating} a method for automatic detection of packet features for OS fingerprinting is presented. Other authors use network flow analysis for OS fingerprinting \cite{mossel2010passive, jirsik2014identifying}. \cite{trimananda2020packet} use packet-level signatures to detect the typical behavior of smart home devices.
%\medskip
%\noindent
%\textbf{Passive Application Fingerprinting:}
Besides OS fingerprinting, there are some approaches for passively fingerprinting specific desktop applications like DNS server software \cite{chitpranee2013towards} based on flows or browsers \cite{garn2019browser} based on TLS handshakes. Both of these approaches focus on one specific type of application and cannot easily be extended to fingerprinting applications in general. Additional work on application fingerprinting based on TLS \cite{anderson2019tls, anderson2020accurate} was done by creating a large knowledge base containing TLS fingerprints together with contextual data like processes, operating systems and IP addresses to train a classifier that is able to detect process families and malware.

Other work focuses on fingerprinting of smartphone apps. \cite{taylor2016appscanner} develop a system to fingerprint mobile apps based on flows of encrypted traffic produced by them. \cite{miskovic2015appprint} use key-value pairs in URLs, HTTP header fields, and traffic flows to automatically generate fingerprints for mobile apps. In \cite{dai2013networkprofiler} the identification of Android apps based on HTTP traffic is investigated. \cite{van2020flowprint} developed a semi-supervised approach to fingerprint apps based on encrypted traffic, which allows to fingerprint previously unseen apps.

DONUT allows for rule-based fingerprinting of applications based on passively monitored DNS traffic. The rules of DONUT can be automatically extracted from application-labeled network traffic generated by our novel network traffic labeling tool ATLAS, which makes our pipeline for application fingerprinting scalable.

%\medskip
%\noindent
%\textbf{Active Fingerprinting:}
Active approaches for fingerprinting operating systems include \cite{arkin2002remote}, where ICMP is used and \cite{greenwald2007toward}, which is based on TCP SYN packets. More generally, network scanning and mapping tools such as \cite{nmap, prtg, advip} allow for the detection of the components within a network as well as the connections between them and thereby determine the current topology of the network and gather various types of information on the devices running in the network including the operating systems they run on and applications installed on them. These tools are typically based on actively scanning the network in combination with deep packet inspection. As such, active approaches are more invasive than the passive DNS-monitoring-only approach taken by DONUT, which does not require any active network access. 

%In contrast to passive fingerprinting, Nmap \cite{nmap} is used for actively scanning network. It can be used to actively probe hosts to reveal specific behavior of different operating systems.

%\medskip
%\noindent
%\textbf{DNS and Privacy:}
The privacy leakage generated by DNS traffic has been under much discussion in recent years \cite{dnsprivacy} and has led to novel suggestions to enhance DNS to offer more privacy (e.g \cite{federrath2011privacy,lu2009ppdns,wachs2014censorship}) or to replace DNS by a more privacy-preserving alternative. It has been shown that DNS traffic allows for a re-identification of individual users \cite{kim2015you}. However, how much information is revealed about the network itself by the DNS traffic generated by the devices in the network is less well studied so far. DONUT contributes to this discussion as it demonstrates that passive monitoring of unprotected DNS traffic at least reveals what software is installed on devices in a network to everyone with access to the DNS traffic. Note that even if DNS traffic is protected by DoH or DoT, the DoH or DoT server may still infer this information as it has access to the full DNS traffic.

\section{Setting and Tool Design}
\label{chap:donut}

Before presenting our approach for data labeling and rule generation, we describe the setting in which DONUT is supposed to be used, its detailed design, and provide an overview on how it is implemented. We start with the setting, then provide a high-level description of the overall structure of the tool, describe its main modules and explain how they interoperate.

\label{sec:setting}
\subsection{Deployment setting}
We start with a description of the deployment settings which DONUT is designed for. The general goal is to capture and analyze DNS traffic (including the IP addresses of the clients initiating the traffic) in order to draw conclusions about the software that is running on each device in a network. This information is mainly useful for network administrators, e.g., to detect unwanted software in Bring Your Own Device (BYOD)-networks or to provide additional context about hosts when handling security incidents. 

\begin{figure}
\centering
\resizebox{\columnwidth}{!}{ 
\includegraphics[scale=1]{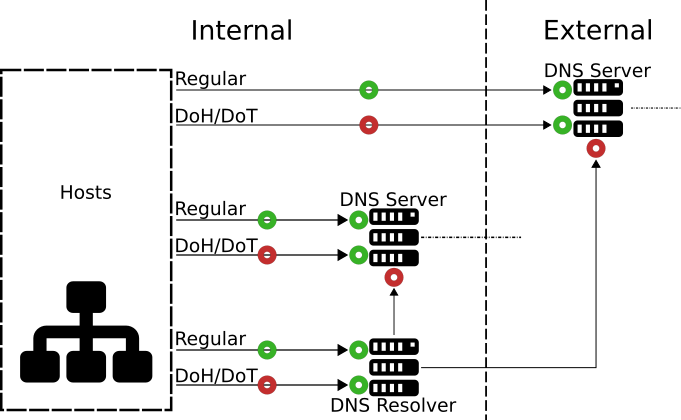}
}
\caption{Possible deployment settings for DONUT: green shapes depict possible deployment positions, red shapes depict positions where DONUT cannot be used.}
\label{fig:setting}
\end{figure}

In general, DONUT monitors unencrypted domains as well as the IP addresses of the querying hosts. Figure \ref{fig:setting} depicts possible deployment positions for DONUT using different combinations of DoH/DoT with an internal or external DNS server. The green shapes indicate positions where DONUT can be deployed, while the red shapes indicate positions where either access to unencrypted domains or access to IP addresses is missing. If an internal DNS server is used, the monitoring can be co-located with the internal DNS server. In this case, the internal DNS server could even apply encryption with DoH/DoT and DONUT could still be used, e.g., by the local network administrator. If an external DNS server is used, DONUT can also be co-located with the server. However, it can only be deployed internally if no DoH/DoT is used, e.g., at a local router. In case a DNS resolver sits between the hosts and the DNS server, e.g., if each subnet uses a separate resolver for caching, DONUT can only be co-located with the resolver because the individual IP addresses are not visible to the internal or external DNS server. 
%Since DONUT analyzes DNS queries, the traffic needs to be captured. This can either be done at a central point in the network, e.g., a router, or at the DNS resolver responsible for the analyzed network. The first method works even if the DNS resolver is external and not controlled by the administrator using DONUT, e.g., in smaller networks without a local DNS resolver where the queries are forwarded to an external server. However, this doesn't work in case encrypted DNS like DoT or DoH is used because DONUT analyzes the queried domains. In this case, DONUT can still be used by capturing the traffic at the network's DNS resolver, assuming it is not external.

We consider DONUT's main use-case to be a tool for analyzing hosts in BYOD networks, especially university networks, where it is common that a central DNS server is used by most of the hosts. This allows to capture the DNS traffic at one place, assuming no additional DNS resolver or NAT configuration is present between the hosts and the DNS server. In such networks it is especially useful to be able to detect running applications because the devices are not centrally managed, which makes it hard to detect unwanted software like crypto-miners and to handle security incidents like malware infections.

\subsection{Design and Implementation of DONUT}
\begin{figure}
\centering
\resizebox{\columnwidth}{!}{ 
\includegraphics[scale=1]{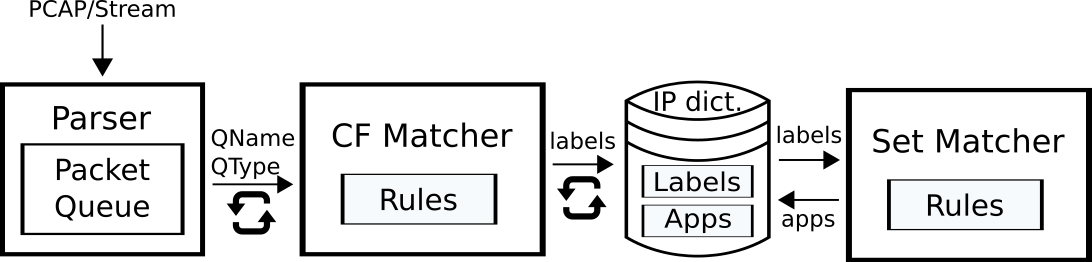}
}
\caption{DONUT's analysis pipeline including its most important modules. The arrows are labeled with ingoing and outgoing data, and the circular arrows depict repeating communication.}
\label{fig:donut}
\end{figure}

%\begin{figure}
%\centering
%\resizebox{0.9\columnwidth}{!}{ 
%\includegraphics[scale=1]{figure/gui.png}
%}
%\caption{Screenshot of DONUT's web-based GUI, showing the distribution of applications %in a network.}
%\label{fig:gui}
%\end{figure}

DONUT takes DNS traffic (pcap files or streams) as input, analyzes it based on pre-configured rules, and after some post-processing, outputs information on the devices that produced the traffic. The input traffic can be captured by tools like TShark\cite{urltshark} and can either take place prior to the analysis or online. The DNS traffic can, e.g., be captured at the router of a network or at the DNS Server directly. The latter allows the usage even if encrypted DNS like DoH is used. Depending on the rules used, DONUT's output includes the applications and services running on each device. Due to its modularity, DONUT has the potential to also produce information about operating systems or NAT configurations. However, in this paper we focus on the core functionality of identifying applications. To make the operation of DONUT as easy as possible, we developed a basic web-based user interface in addition to command-line enabling visual inspection of the results. 
%It allows to edit settings, start the analysis process, access and reprocess the results, and manage previous runs. The DNS traffic can either be uploaded as pcap file, or streamed via the network for online analysis. The results are accessible via various graphs, e.g., for the distribution of applications in the whole network.% as seen in Figure \ref{fig:gui}. 

DONUT uses a rule-based approach and is designed to be as modular and extensible as possible. In this paper, we discuss two types of rules, namely CF-rules and Set-rules. A CF (context free)-rule simply encodes a specific queried domain and adds a corresponding label to all IPs querying that domain. The Set-rules then combine these labels in a post-processing step to produce the final results, i.e., the applications that most likely produced this combination of DNS queries. Both types of rules are generated automatically based on labeled data, which we describe in Section \ref{chap:fingerprint_generation}.

As depicted in Figure \ref{fig:donut}, DONUT is composed of three main modules, and an IP dictionary. The parser extracts the relevant fields from each DNS packet, i.e., IP, domain name, and query type, from the traffic and puts them into a processing queue. The analyzer module, containing the CF-Matcher, pulls the packets successively from the queue, checks whether they match to a CF-rule, and updates DONUT's data structure accordingly. In general, the CF-rules only take information of a single DNS packet into account. More complex rules, e.g., encoding sequences of packets, can be added in the future due to DONUTs modularity. Finally, in the post-processing module, the Set-matcher processes the stored labels according to the Set-rules, and computes the list of identified applications for each IP. The main data structure of DONUT is a dictionary for each IP address active in the network traffic. In the following, we will refer to this data structure as IP dictionary. All information gathered by the Analyzer, i.e., the labels added by the CF-rules, are stored in that data structure. All rules and mappings to application names are stored in configuration files.
%Currently, DONUT uses three configuration files. One for each type of rule, and one additional file for a mapping between internal IDs (which are used in the rules) and the actual application names that are displayed in the output. 

%\subsection{CF-Matcher}
%\label{sec:cfmatcher}

\begin{figure}
\centering
\resizebox{0.8\columnwidth}{!}{ 
\includegraphics[scale=1]{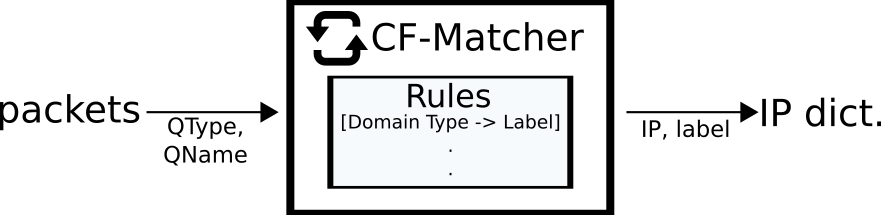}
}
\caption{Structure of the CF-Matcher, which adds labels to the IP dictionary depending on which rule has matched to the host's DNS queries.}%The arrows are labeled with ingoing and outgoing data, and the circular arrow depicts the repeating execution of the CF-Matcher.
\label{fig:cfmatcher}
\end{figure}

\begin{table}
\caption{Entries of the CF-Matcher rules, which are compared to each packet in the pcap file or stream.}
\label{tab:donut_cfentries}
\begin{minipage}{\columnwidth}
\begin{center}
\resizebox{\columnwidth}{!}{ 
\begin{tabular}{ll}
\toprule
\textbf{Entry} & \textbf{Description} \\
\midrule
Domain name & The domain name to match. \\
Query type &  The query type(s) of the packet to match. \\
Label & A list of labels to indicate that this rule has matched. \\
\bottomrule
\end{tabular}
}
\end{center}
\end{minipage}
\end{table}

%\begin{lstlisting}[caption={Example of a CF-rule matching to a domain from the crypto-mining application Nicehash.},label={lst:cfrule},captionpos=b, float]
%#Domain name                      QType  Label
% randomxmonero.eu.nicehash.com	  1;28   nicehash_5
%\end{lstlisting}

The CF-Matcher analyzes domain names of single packets without taking other context into account, as depicted in Figure \ref{fig:cfmatcher}. Applications use DNS to resolve the IP addresses of their remote communication endpoints. Hence, the set of DNS queries sent by an application can already be unique to this application and is thus usable to fingerprint an application without taking additional information into account. 
%As a side note, it is clear that this method can be circumvented by manually querying all domains used in CF rules to obfuscate the actually running applications. However, our goal is not to create a tool capable for a scenario where users actively try to hide or manipulate their traffic, but rather to build a tool to gain a quick overview of applications running in a network, e.g., to detect outdated or unwanted software as a network administrator.
Each CF-rule contains the entries listed in table \ref{tab:donut_cfentries}. The Domain name and the query type are compared with all packets from the pcap file or stream by the analyzer. If the entries match to a packet, the list of labels, which is usually just one, is added to the dictionary of the corresponding IP address. In case multiple applications query the same domain, multiple labels for different applications can be added by one CF-rule.

%\subsection{Set-Matcher}
%\label{sec:setmatcher}

The Set-Matcher operates as a post-processing step once the analysis of the pcap file or stream is finished. Additionally, it can be configured to be executed regularly during analysis to compute preliminary results, e.g., every few minutes or after a certain amount of packets. Its task is to combine all information gathered during the analysis and compute corresponding implications, i.e., which applications are running, based on the labels added by the CF-Matcher. Each application is represented by one Set-rule, containing the entries listed in Table \ref{tab:donut_setentries}. Since the Set-Matcher is not part of the analyzer, its rules do not contain packet entries as prerequisites. This implies, that the Set-rules can also be changed after the packet analysis is already finished, as long as they only contain labels of CF-rules that were used during the analysis. This allows for a later reconfiguration of Set-rules as needed, e.g., changing of rule parameters to improve false positives or false negatives. 

The prerequisites of Set-rules are a combination of required labels, optional labels, and a threshold. All required labels must be present for a rule to match. Of the optional labels, at least the stated minimum number of labels need to be present, in addition to the required labels. The motivation for this differentiation is that applications often connect to a (potentially small) set of communication endpoints, and therefore query the corresponding domains on each start or at regular intervals during runtime. Other DNS queries might only be triggered when a certain function of the application is used, e.g., during updating or an actively initiated video call. The idea is to configure domains observed in most runs or instances during the rule generation process as required and all domains queried during only a few runs as optional. The goal of optional labels, which are chosen less conservative as required labels, is to reduce the number of false positives, especially in case the list of required labels is short. The minimum number of optional labels can be used to adjust the risk of false negatives. Another use-case for the concept of optional labels is to be able to include domains that are also queried by other applications or are likely to be queried by a user, e.g., while browsing the web. In this case, the minimum number is used as a threshold to reduce false positives while the required labels ensure that, e.g., domains likely to be queried by a user alone are not sufficient for a Set-rule to be triggered. The process of generating these rules is automated and described in more detail in Section \ref{chap:fingerprint_generation}.

\begin{figure}
\centering
\resizebox{0.8\columnwidth}{!}{ 
\includegraphics[scale=1]{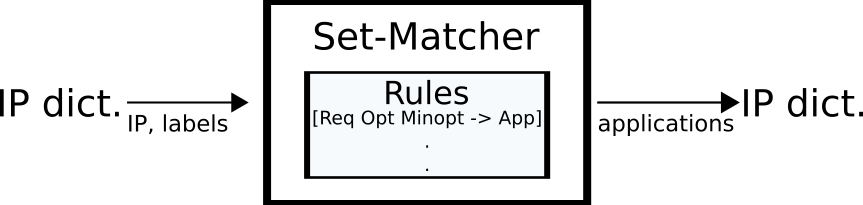}
}
\caption{Structure of the Set-Matcher, which combines the labels added by the CF-Matcher in order to conclude which applications are running on a host.}
\label{fig:setmatcher}
\end{figure}

\begin{table}
\centering
\caption{Entries of the Set-Matcher rules, which combine the information gathered in the analysis step.}
\label{tab:donut_setentries}
\begin{minipage}{\columnwidth}
\begin{center}
\resizebox{\columnwidth}{!}{ 
\begin{tabular}{ll}
\toprule
\textbf{Entry} & \textbf{Description} \\
\midrule
Required labels & Labels required to be present. \\
Optional labels & Optional labels. \\
Min. optional & Minimum number of required optional labels. \\
%Forbidden labels & Labels that must not be present. \\
ID implication & ID of implied application. \\
\bottomrule
\end{tabular}
}
\end{center}
\end{minipage}
\end{table}

%\begin{lstlisting}[caption={Example of a CF-rule combining multiple labels to imply the crypto-mining application Nicehash, referenced by applicaion ID 107.},label={lst:cfrule},captionpos=b, float]
%#Required  Optional                        Min  Implication
%nicehash_5 nicehash_1;nicehash_2;nicehash6 2    107
%\end{lstlisting}

\subsection{Limitations}
\label{sec:limitations}
We presented DONUT as a tool to fingerprint applications based on DNS traffic. For obvious reasons, this works only for applications that use DNS in a way that is distinguishable from other applications. However, even applications that can operate completely offline often produce DNS traffic, e.g., when searching for updates. If two applications, e.g., from the same vendor, query the same set of DNS queries, it is not possible to distinguish them without additionally analyzing other parameters like timing or order of queries. Also, it is clear that our fingerprinting method can be circumvented by manually querying all domains used in CF rules to obfuscate the actually running applications. However, our goal is not to create a tool capable for a scenario where users actively try to hide or manipulate their traffic, but rather to build a tool to gain a quick overview of applications running in a network, e.g., to detect unwanted software as a network administrator. As discussed in Section \ref{sec:setting}, DONUT can still be used if the DNS traffic is encrypted, as long as it is possible to directly capture the traffic on the central DNS server of the network.
\section{Rule Generation}
\label{chap:fingerprint_generation}
In this section, we describe our methodology for the automated generation of rules for DONUT. We present our tool ATLAS used for traffic labeling, describe the procedure for data generation for a selection of applications to test our approach, and present the process of rule extraction in detail.

%\subsection{Overview}
%\label{sec:overview}

\begin{figure}
\centering
\resizebox{\columnwidth}{!}{ 
\includegraphics[scale=1]{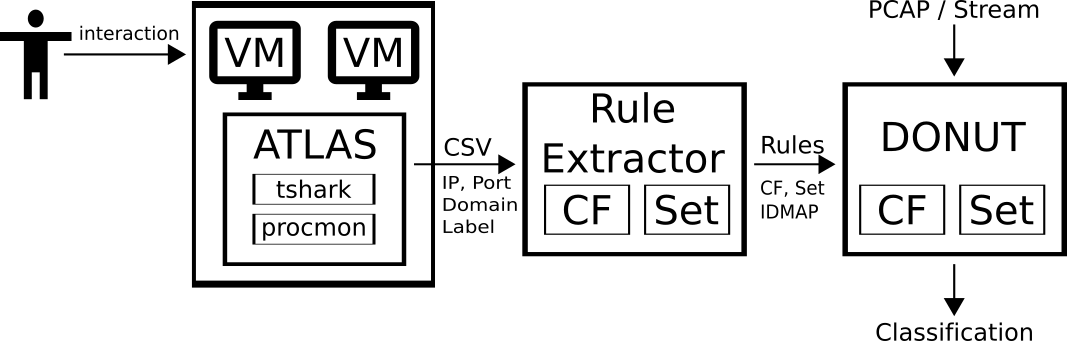}
}
\caption{Complete fingerprinting pipeline, including data generation with ATLAS, rule extraction, and analysis with DONUT.}
\label{fig:pipeline}
\end{figure}

The process for the generation of rules is composed of two parts, namely (1) data generation and labeling, and (2) rule extraction. The resulting rules are then used by DONUT for application fingerprinting. The complete pipeline is depicted in Figure \ref{fig:pipeline}. For labeling, we developed the Automated Traffic Labeling Software (ATLAS), which is used to label network packets by the process name that created them with ground truth accuracy. We used ATLAS on multiple virtual machines, where we installed and interacted with a number of applications to generate the DNS traffic for labeling. Based on this data, we used the Rule Extractor to generate rules for DONUT.

\subsection{Labeling}
\label{sec:datageneration}

In the following, we describe ATLAS in more detail. It monitors network traffic and system events and currently it supports Windows 10. ATLAS includes a simple GUI to start and stop the monitoring and to adjust settings, e.g., the network protocols which should be monitored. The way network packets are handled differs between operating systems, however, for future work we plan to add support for other operating systems like Linux or MacOS. ATLAS is able to label packets of all network protocols with the corresponding application, of which we only need DNS for this paper. The tool consists of mainly three components: event logging, packet capturing, and packet labeling. 

The event logging component uses the Windows Process Monitor (procmon), which monitors system events in real-time, e.g., file system access or process activity. It logs all events occurring on a machine together with the application that caused the event. These events are then filtered for network events, and parsed, using wtrace\cite{urlwtrace}. In addition to procmon, the logging component polls tasklist.exe periodically to receive the process ID of each process. Also, it activates and parses the DNS event log of Windows, which is used to track the process IDs of applications calling the system's DNS resolver. For each event, the timestamp, process name and ID, source and destination IP, as well as source and destination port is forwarded to the labeling component via a queue. Simultaneously, the capturing component uses TShark to capture network traffic in real-time. For each packet, the frame number, timestamp, source and destination IP, as well as source and destination port, are parsed and forwarded to the labeling component via another queue.

The labeling component receives system events and network packets and attempts to match each event to a packet. This is done by comparing the IP addresses and ports of an event to the packets from the queue. If all parameters match and the timestamps of event and packet are within a small time frame, the packet with the smallest time-delta is labeled with the corresponding process name, and packet as well as event are deleted from their queue. If the timestamps of event and packet are slightly off, it is possible that the event used for labeling the packet is not the event that caused the exact packet, e.g., if the corresponding process generates many packets in fast succession. However, because the combination of IP addresses and ports can only be used by one process at a time, this still results in correctly labeled packets. Considering DNS, applications use the DNS resolver of the operating system to resolve a URL. Hence, since procmon reports which process triggered a network event, this would result in all DNS packets being labeled with the DNS resolver process. To solve this, the DNS packets are compared to the DNS events also collected by the logging component, the events with the smallest time delta are matched, and the process ID in the DNS event is replaced by the process name for the label, using the information collected from tasklist.exe. In order to label packets without a preceding system event, e.g., follow up packets of a TCP connection, the tool additionally tracks all connections that were successfully labeled at least once. Packets that cannot be matched directly with a system event are compared to known connections, defined by source and destination IP addresses and ports, in order to assign the correct label.

The output of ATLAS consists of the pcap file captured by TShark and a csv file which contains the frame number for each packet from the pcap, timestamp, source and destination ports and IP, process name and id, and the domain in case of DNS packets. In summary, this results in a labeling coverage of almost 100\%, which would not be possible by simply polling open network connections because of short-lived connections such as DNS. The only packets that cannot be labeled are the ones without a preceding system event, that belong to a connection, which was opened before the labeling process started, and closed after the labeling stopped, e.g., TCP keepalive packets.

\subsection{Data Generation and Application Selection}
\label{sec:softwareselection}
Next, we describe the process of generating network traffic for a selection of applications. We use ATLAS to label this traffic in order to automatically extract rules that can be used for fingerprinting with DONUT. We briefly describe our data generation setup, motivate the selection of applications, and outline how we interacted with the applications in order to generate DNS traffic. We used Proxmox Virtual Environment\cite{urlproxmox} where we set up virtual machines with Windows 10. On each VM we installed ATLAS as well as the applications we want to generate labeled network traffic for. Each application was installed on multiple VMs and interacted with repeatedly in order to generate as much DNS traffic as possible.

We selected the applications based on two categories. First, widely used applications that are installed on most office computers, and second, applications that might be unwanted in some networks due to organizational policies. We look at widespread applications in order to test the generated rules on an unlabeled real world dataset. With the second category, we aim to demonstrate another practical use case of our fingerprinting approach, i.e., finding unwanted software in a network as an administrator. For the "widespread" category, we installed the browsers Chrome, Firefox, and Edge; Microsoft Office including Word, Excel, Powerpoint and Outlook; Skype and Zoom as video conference software; and Sophos as an antivirus solution. The latter was selected because it is widespread in the RWTH university network for which we obtained a large real-world dataset in order to test DONUT. As potentially unwanted software we installed crypto-miners, namely Easyminer and Nicehash; cloud storage software OneDrive and Dropbox due to privacy concerns; the gaming platform Steam; and Teamviewer because of its potential for unauthorized access. Additionally, all traffic from Windows was captured to generate rules for the operating system processes itself.

We manually interacted with all applications, started and stopped each applications repeatedly, and passively ran the applications for multiple hours. For example, we created accounts for each application that require a login, e.g., Outlook, Skype, and Zoom; used the application's main functionality, e.g., browsing, video calls, crypto mining, or virus scans; and triggered updates. This most likely does not result in the complete spectrum of possible DNS traffic for each software, however, it generally produced a sufficient amount of traffic to be used for rule extraction. In total, we generated and labeled network traffic for 14 applications as well as Windows 10 resulting in 989.922 packets in total, of which 61.407 are labeled DNS packets containing 10.096 different queries. Note, that 8.514 of the 10.096 unique domains were queried by Sophos alone, because Sophos excessively uses DNS to lookup IP addresses and presumably file hashes during malware scans.

%For future work, we plan install ATLAS and generate data on real-world machines. This would allow for collecting more realistic data, but it is also complicated due to the privacy critical nature of this data, especially DNS. However, the data from VMs is sufficient for a proof-of-concept of our mostly automated pipeline for application fingerprinting. Additionally, we will look into ways to (partially) automate the interaction with applications, to make the data generation process more scalable.

\subsection{Rule Extraction}
\label{sec:ruleextraction}
After covering the process for labeling and data generation, we now describe our approach of automated rule extraction for DONUT's fingerprinting capabilities. The Rule-Extractor consists of three main components, namely a parser, the CF-Rule generator, and the Set-Rule generator. It also implements filtering functionality for the data and rules. For each component, several parameters can be adjusted to configure the logic for rule extraction.

The parser takes a number of csv files generated by ATLAS as input and extracts all labeled DNS packets. The labels are compared against a whitelist filter, which is the only part that requires manual input. The whitelist specifies all processes that should be considered for rule extraction, which is necessary for two reasons. First, to exclude irrelevant processes, e.g., sub processes from the operating system that are not of interest, and second, to combine multiple process names into one application name. For example, we combined the most common present processes of Windows 10 (e.g., \textit{svchost.exe} and \textit{sihclient.exe}) under one label identifying Windows 10. As another example, the application Teamviewer consists of two processes, namely \textit{temviewer.exe} and \textit{teamviewer\_service.exe}, which we also combined under one label. All processes that are not whitelisted are combined under the label \textit{other}, in order to not loose the information, that some domains queried by a whitelisted process might also be queried by another one not on the whitelist. 
%In general, it is also possible to skip the whitelisting step and extract rules for each process independently. However, for example in the case of Teamviewer, both processes have an overlapping set of queried domains, which means less unique domains for each process, leading to less uniquely identifying domains to generate rules from. For future work, it may be possible to link such processes together automatically, e.g., by searching the installation folder of an application for all executables. 
Another filtering step of the parser is comparing all domains to the Tranco list \cite{pochat2018tranco}, containing the most popular websites. We do this as an easy way to filter domains that are likely to be queried by a user, e.g., through a browser, in order to lower false positives. Additionally, we filter queries containing the local domain of the network our data generation VMs operate in, to prevent that such a domain ends up as required in a Set-rule, which would lead to many false negatives in a different network environment.

The CF-rule generator creates a preliminary list of CF-rules based on the parsed information. CF-rules only imply labels instead of applications directly and some of them are not used in any Set-rule. In this case, the corresponding CF-rule can be removed. However, to allow for dynamic re-configuration of Set-rules after the analysis, it makes sense to still use all available CF-rules to increase the number of usable labels for Set-rules. For each whitelisted application, one CF-rule with a unique label for each queried domain of this application is generated. The Set-rule generator creates one rule for each application based on the labels from CF-rules. Which domains are used for the Set-rule of one application depends on the three criteria, shown in Table \ref{tab:set_criteria}.

\begin{table}[!t]
\caption{Criteria used to decide which domain labels are used in a Set-rule for one application.}
\label{tab:set_criteria}
\begin{minipage}{\columnwidth}
\begin{center}
\resizebox{\columnwidth}{!}{ 
\begin{tabular}{lp{.68\columnwidth}}
\toprule
\textbf{Criterion} & \textbf{Description} \\
\midrule
isUnique & Boolean value, specifying whether a domain was queried by only one or multiple applications. \\
occurence & Value between 0 and 1, depicting the portion of application instances in the dataset where the domain was queried relative to the total amount of instances of this application. \\
medianInterval & Median interval between queries of the same domain by one application. \\
\bottomrule
\end{tabular}
}
\end{center}
\end{minipage}
\end{table}

The boolean value \textit{isUnique} indicates whether a domain was queried by one or multiple applications in the dataset. In the latter case, the domain's label is disregarded as required label in the Set-rule. However, non-unique queries can be used for optional labels, since they only become relevant if all required labels of an application are present as well.  

\textit{Occurrence} indicates, how many different instances of an application in the dataset queried the domain. In case of, e.g., a browser, most domains resulting from browsing the web will most likely have a lower \textit{occurence} value, if the dataset contains enough instances of the browser with different browsing behavior. This is because not all users browse the same websites, e.g., \textit{www.scholar.google.com} will only appear in a fraction of browser instances recorded. On the other hand, domains that are queried by the browser without user interaction, e.g., checking for updates, will have a higher \textit{occurence} value. In the Set-rule generator, this value is, e.g., used to decide whether a label should be \textit{required} or \textit{optional}, as described in Chapter \ref{chap:donut}.

The third criterion, \textit{medianInterval}, represents how often a specific domain is queried by one application instance, defined by the median interval between two repetitions of the same query. We use median instead of average to reduce the influence of time frames where an application is not running in the dataset. This value is used to prioritize the domains, i.e., labels, that are used for a Set-rule. The higher the frequency of a domain, the higher the chance that a domain is queried by an application during the runtime of DONUT. 

\begin{table}[!t]
\caption{Configurable parameters used for Set-rule generation based on the criteria listed in Table \ref{tab:set_criteria}.}
\label{tab:set_parameters}
\begin{minipage}{\columnwidth}
\begin{center}
\resizebox{\columnwidth}{!}{ 
\begin{tabular}{lp{.68\columnwidth}}
\toprule
\textbf{Criterion} & \textbf{Description} \\
\midrule
maxReqLabels & Max. \textit{required} labels in a Set-rule. \\
maxOptLabels & Max. \textit{optional} labels in a Set-rule. \\
minReqOccurrence & Min. occurrence value for \textit{required} labels. \\
minOptOccurrence & Min. occurrence value for \textit{optional} labels. \\
minOptProbability & Desired probability that the minimum number of optional labels (see Table \ref{tab:donut_setentries}) are present, based on the occurence value. \\
maxReqMedian & Max. median interval for \textit{required} labels. \\
maxOptMedian & Max. median interval for \textit{optional} labels. \\
allowOptNonunique & If True, domains queried by multiple applications can be used for optional labels. \\
%minDomainRepetition & Minimum number of repetitions of a single domain for it to be considered for a Set-rule.\\
\bottomrule
\end{tabular}
}
\end{center}
\end{minipage}
\end{table}

The three criteria are computed for each domain of an application. Defined by configurable parameters, listed in Table \ref{tab:set_parameters}, the Set-rule generation works as follows. For an application, each label, corresponding to a domain of the application, is classified as a candidate for \textit{required}, \textit{optional}, or nothing. If a label has an \textit{occurrence} value larger than \textit{minReqOccurrence} and a \textit{medianInterval} value lower than \textit{maxReqMedian}, it is considered a candidate to be a \textit{required} label. The \textit{minReqOccurrence} parameter ensures that the domain was queried by a large enough portion of application instances, while \textit{maxReqMedian} ensures that the domain is expected to be queried regularly. This ensures that each \textit{required} domain is expected to be queried by future instances of the application and that it is expected to occur in a small enough time frame. The same categorization is done for \textit{optional} labels. However, with less restricted parameters, since only  a portion of optional labels need to be present for a Set-rule to trigger.

Once the candidates for \textit{required} labels are selected, the list is sorted by their median interval in ascending order to prioritize domains queried more frequently, such that the resulting Set-rule can trigger as early as possible on live monitored traffic. If the list contains more than \textit{maxReqLabels}, the excess candidates are added to the list of optional candidates and the most frequent ones are selected as required labels. Finally, the list of optional candidates is sorted by the median interval as well, and at most \textit{maxOptLabels} are selected.

%The maximum number of labels is capped to not end up with dozens of labels, and therefore rules, for each application, which negatively impacts the analysis performance of DONUT.
For required labels, we only allow domains that were queried by exactly one application in the dataset. For optional labels, the parameter \textit{allowOptNonunique} controls whether domains queried by multiple applications are allowed. However, even if only unique domains queried by exactly one application in the dataset are considered, there is a chance that a domain is also queried by an application which is not in the dataset, or that a domain is queried by a user instead of an application. By limiting the number of domains in rules and prioritizing domains that are queried often and by most application instances in the dataset, the chance of including a user dependent domain in the rules is limited. Any domain can be queried by a user manually, however, the goal of this pipeline is not to be robust against active avoidance, e.g., if someone manually queries domains of all DONUT rules.

%As a baseline, we limit the number of required labels to 5, with a minimum occurrence value of 1 to reduce false positives. These parameters ensure a high confidence that \textit{required} labels are representative for an application. For optional labels, we limit the number to 10, to have more potential candidates because not all labels need to be present, while choosing a minimum occurrence value of 0.5, such that all domains were queried by the majority of hosts in the dataset. The optional labels are chosen less conservatively, in order to consider more labels in each Set-rule to reduce the number of false positives, especially in case less then 5 required labels were chosen. We also allow non-unique domains to be used for optional labels. We discuss the impact of changing \textit{maxReqLabels} and \textit{maxOptLabels} in \ref{chap:evaluation}
%In order to be able to compute meaningful median interval values, we set \textit{minDomainRepetition} to 4 as a baseline, such that at least three intervals can be computed.

The number of \textit{optional} labels that need to be present in order for a Set-rule to trigger is calculated based on \textit{minOptProbability}, representing the desired chance that this number of queries is actually present based on the occurrence value. We interpret the occurrence value as the chance that a domain is queried by an application instance, as it represents the portion of instances in the dataset which queried the domain. The minimum number of optional labels is then equal to the maximum number \textbf{n}, such that the chance of at least \textbf{n} domains being queried is greater than \textit{minOptProbability}. %Choosing a high probability can result in more false positives, because the minimum number of optional labels tends to be lower, while a low probability can result in more false negatives. As a baseline, we chose a probability of 0.8. However, we discuss the impact of this value in \ref{chap:evaluation}.

%Note, that these parameters were chosen with our generated dataset in mind. If data from more diverse environments is used, it might be better to choose less conservative parameters to ensure that enough label candidates exist, because the intersection of domains queried by instances of an application might become smaller, the more instances are considered.

In Section \ref{chap:evaluation}, we select a baseline set of parameters and discuss the impact of each parameter on the accuracy measured on an artificial validation dataset with known ground truth. The selected set of parameters is then used to test DONUT on a real-world dataset with unknown ground-truth.

%\todo[inline]{TODO (ca. 1 column, either here or in eval) present some statistics/characteristics of resulting rules. e.g. number of labels per application, ratios of required and optional labels, characteristics of occurrence}

\section{Evaluation}
\label{chap:evaluation}

In the following, we present the evaluation of performance and functionality of our pipeline. We first introduce the datasets used for evaluation. Next, we discuss performance and resource consumption of ATLAS, rule extraction, and DONUT. We analyze the impact of different parameters for Set-rule extraction and evaluate the correctness of the analysis on an artificial validation dataset with known ground truth. Finally, we discuss the output of DONUT on a large real world dataset for which the ground truth is unknown.

\subsection{Evaluation Datasets}

\begin{table}[!t]
\caption{Evaluation datasets containing DNS traffic of real-world and artificial networks.}
\label{tab:datasets}
\begin{minipage}{\columnwidth}
\begin{center}
\resizebox{\columnwidth}{!}{ 
\begin{tabular}{llll}
\toprule
\textbf{Dataset} & \textbf{Source} & \textbf{Size} & \textbf{\# Hosts} \\
\midrule
Validation set & Virtual machines & 9.273 DNS packets & 4 \\
Performance set & RWTH university network & 1M DNS packets & 23.455 \\
Real-world set & RWTH university network & 216M DNS packets & 43.961 \\ 
\bottomrule
\end{tabular}
}
\end{center}
\end{minipage}
\end{table}

We use the three datasets for evaluation of performance and functionality of our pipeline, listed in Table \ref{tab:datasets}. The validation set was manually created using four virtual machines in the same environment as descried in Section \ref{chap:fingerprint_generation}. We installed the same applications that we created rules for on three virtual machines in random combinations, and interacted with them in a similar way. Additionally, we installed all applications on a fourth VM, but used them for only one minute before closing them again, in order to produce edge cases for our evaluation. Secondly, we used a large real-world dataset captured roughly one year before writing this paper on the edge routers of the RWTH university network. The dataset contains only DNS traffic and all IP addresses were anonymized with the tool \textit{capsan}\footnote{capsan: a pcap anonymizer, from https://github.com/jsiwek/capsan} before it was sent to us. We split this into a small set for performance evaluation and a large set for functional testing. The latter contains 7.5 hours of traffic produced by 43.961 hosts. We have no ground-truth for this dataset. While this does not impact the performance evaluation, we can only discuss the plausibility of the analysis results on the real-world dataset and cannot evaluate false positives or false negatives.

\subsection{Performance}
Next, we present the performance evaluation of ATLAS, rule-extraction, and DONUT. The measurements were done on a virtual machine running on a server with an AMD Epyc 7702P processor, with an allocation of 8 cores and 16GB memory. ATLAS managed to label everything in real-time with a small delay, because matching packets to events are searched for in a small time window. This was achieved while labeling all network packets, however, peaks while buffering videos for example, also caused the buffer of unlabeled packets to rise temporarily. During labeling, CPU utilization ranged between 5\% and 15\%, depending on the amount of packets in the buffer. When only labeling DNS packets, which is sufficient in our context, no delay occured because DNS corresponds to only a fraction of the overall produced packets. Hence, we didn't focus on further optimizing the tool's performance.

The rule extraction performance largely depends on the maximum number of optional labels for Set-rules, because of the exponential complexity of the calculation for \textit{minOptProbability} described in Section \ref{chap:fingerprint_generation}. However, even with 20 optional labels for the number of applications we generated data for, the rule extraction took less than 30 seconds. Note, that we did not focus on optimizing performance for ATLAS and rule extraction yet, as their performance is sufficient for our current experiments. 

For DONUT, we measured the number of packets that can be analyzed per second in order to determine whether traffic of large networks can be analyzed in real-time. DONUT's performance depends on TShark, which is used as parser, and on the CF-Matcher. Performance of the latter does not depend on the number of rules, hence it scales well even with a large number of applications. On average, the analysis on our performance dataset took 67 seconds, corresponding to almost 15.000 packets analyzed per second. We measured the performance with a varying amount of CF-rules. However, the runtime stayed constant which makes DONUT scalable to support rules for many applications. The real-world RWTH dataset contains 216M DNS queries with 44.000 active hosts over the course of 7.5 hours which corresponds to 8.000 packets per second on average. Even when the amount of traffic peaks temporarily, DONUT's performance is sufficient with a maximum of a few seconds delay when traffic exceeds the capabilities and packets need to be buffered. To scale our approach for even larger networks, multiple instances of DONUT can run simultaneously, e.g., one for each subnet. For our evaluation, one instance was sufficient.

\subsection{Rule-extraction Parameters}

\begin{figure}
\centering
\resizebox{\columnwidth}{!}{ 
\includegraphics[scale=1]{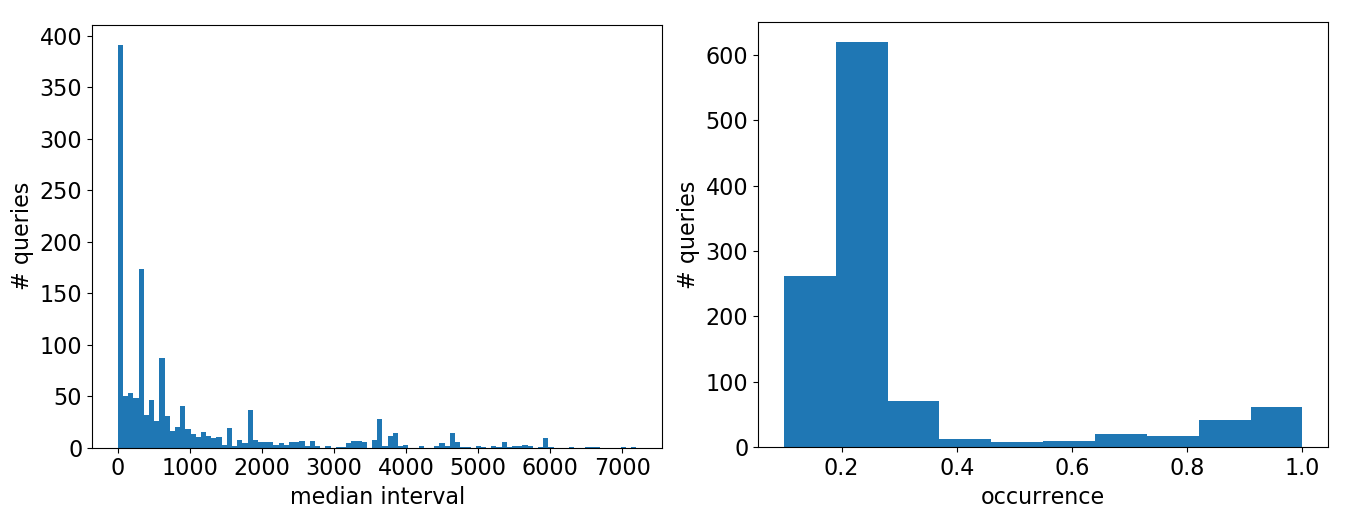}
}
\caption{Histogram of median intervals between queries of the same domain by one application (left), and occurrence values of all domains with a median interval of 1400 and less (right).}
\label{fig:medianocc}
\end{figure}

Now we will discuss the impact of the parameters for Set-rule extraction, listed in Table \ref{tab:set_parameters}, on the classification, and determine a set of parameters to test DONUT on the real-world dataset. We start with a baseline for the parameters and compute the false positives and false negatives for different amounts of required and optional labels for the Set-rules. Afterwards, we discuss the impact of the parameters on the classification of our validation dataset.

For \textit{minOptProbability}, we set a value of 0.5. Figure \ref{fig:medianocc} (left) shows the distribution of \textit{medianIntervals} among all domains. Most domains have a median interval between 1 and 1400, with occasional peaks above 1400, hence we set \textit{maxReqMedian} and \textit{maxOptMedian} to 1400. Since labels are prioritized by low median values, we do not set a lower maximum for required labels. Figure \ref{fig:medianocc} (right) shows the distribution of occurrence values for all domains with a median interval below 1400. Many domains have an occurrence value below 0.4, and another peak can be seen at a value of 1.0. We set \textit{minReqOccurrence} to 1.0, such that only domains that were queried by all application instances are required to be present. For \textit{minOptOccurrence}, we chose a value of 0.6, in order to require that more than half of all application instances queried the domain during data generation. We set \textit{allowOptNonunique} to true, in order to allow for more potential labels, especially for applications with a low total number of queried domains. Since we have no way to determine good values for \textit{maxReqLabels} and \textit{maxOptLabels}, we evaluate the classification results of our validation set with all combinations of 0-10 required and optional labels.

\subsection{Validation Set Results}

\begin{figure}
\centering
\resizebox{0.8\columnwidth}{!}{ 
\includegraphics[scale=1]{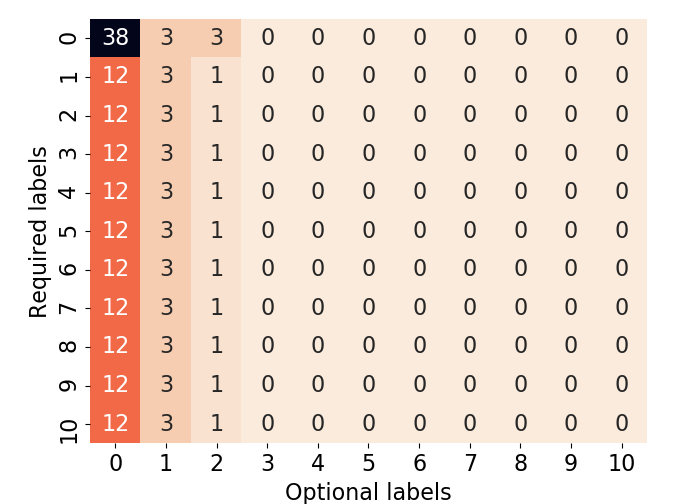}
}
\caption{Heatmap of false positives on the validation datset for all combinations of 0-10 required and optional labels in Set-rules.}
\label{fig:validation_fp}
\end{figure}

\begin{figure}
\centering
\resizebox{\columnwidth}{!}{ 
\includegraphics[scale=1]{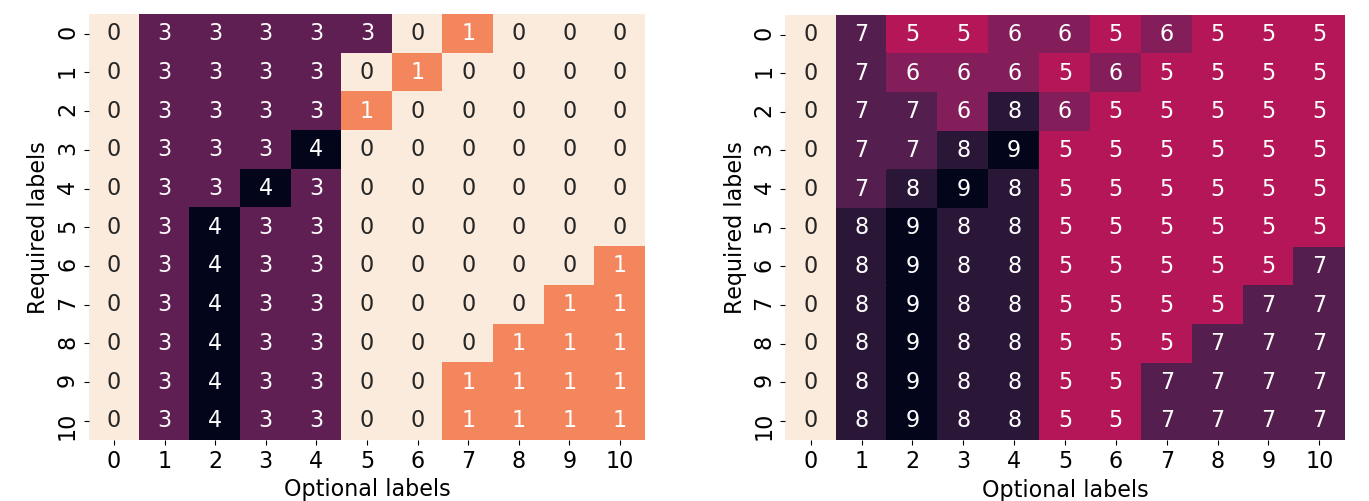}
}
\caption{Heatmap of false negatives on the validation dataset, without short application run times (left) and with short application runtimes (right), for all combinations of 0-10 required and optional labels in Set-rules.}
\label{fig:validation_fn}
\end{figure}

We used DONUT to analyze the validation dataset using rules extracted with the described parameters. Figure \ref{fig:validation_fp} shows the false positives with different combinations of \textit{maxReqLabels} and \textit{maxOptLabels}. It shows that the number of required labels in Set-rules has almost no impact on false positives, and for 3 or more optional labels we observe zero false positives on the validation set. Figure \ref{fig:validation_fn} shows the false negatives for different number of labels, once without the host where each application was only ran for one minute (left), and once including this host (right). Without these edge cases, roughly half of all combinations result in zero false negatives. With the edge cases, the lowest number of false negatives is five for all combinations, excluding the ones with zero optional labels. The false negatives are caused by not detecting Dropbox, Zoom, Teamviewer, Sophos, and Windows 10. Apparently, for these applications and the operating system itself, the amount of DNS traffic during the short time frame was not enough for detecting them. However, we successfully detected the remaining 9 applications. We conclude that our fingerprinting works well on the validation set, with zero false positives and only a few negatives for applications monitored for a short period of time only. For the real-world dataset, we decide to use a maximum of 4 required labels and 7 optional labels for our Set-rules, to minimize the amount of false negatives.

Next, we describe how changing the rule-extraction parameters impacts the classification. For values of \textit{minReqOccurrence} below 1.0 we observed an increment in false negatives, presumably because some of the additional required labels are not present in the validation set. For \textit{minOptOccurrence}, values below 0.5 caused more false negatives, especially for the detection of Windows 10, while values above 0.65 resulted in zero labels for Sophos. The reason for this is that Sophos mostly queries domains containing IPs or file hashes, each with low occurrence values, while we observed no domain that was queried by all Sophos instances. With values above 0.5 for \textit{minOptProbability}, we observed an increasing amount of false negatives, especially for Windows 10. We observed no meaningful impact when lowering \textit{maxReqMedian} besides an increment in false negatives below 120. Dropping \textit{maxOptMedian} below 1200, resulted in an increasing amount of false negatives for Windows 10. With \textit{allowOptNonunique} set to false, we also observed more false negatives. The same holds for not excluding domains from the Tranco list. In general, changing the parameters mostly impacted the amount of false negatives, while the amount of false positives only increased when changing the parameters drastically.

\begin{figure}
\centering
\resizebox{\columnwidth}{!}{ 
\includegraphics[scale=1]{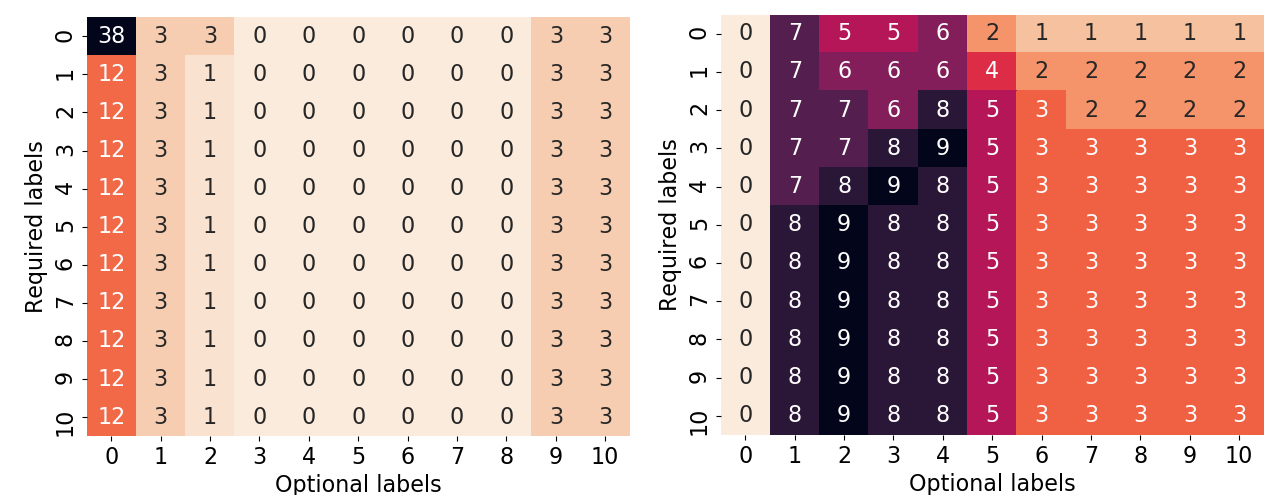}
}
\caption{Heatmap of false positives (left) and false negatives (right) on the validation datset with a minOpt value capped to 4, for all combinations of 0-10 required and optional labels in Set-rules.}
\label{fig:fp_fn_capped}
\end{figure}

When manually inspecting the extracted rules, we noticed that false negatives mostly occurred for applications where the \textit{minOpt} value is equal to or slightly lower than the total amount of optional labels. This could be an indication for overfitting, because we interacted with the applications using repeatable patterns, instead of collecting data from real-world machines. Therefore, we artificially capped the value for \textit{minOpt} as an attempt to further reduce false negatives. With a maximum value of 4, we still end up with low false positives while false negatives are reduced, as seen in Figure \ref{fig:fp_fn_capped}. We achieve the best results with zero required labels and 6-8 optional labels.

\subsection{Real-world Dataset Results}

\begin{table}[!t]
\caption{Fingerprinting results of DONUT applied to the real-world dataset, with and without capped \textit{minOpt}.}
\label{tab:realworld_results}
\begin{minipage}{\columnwidth}
\begin{center}
\resizebox{\columnwidth}{!}{ 
\begin{tabular}{lll||lll}
\toprule
\textbf{Software} & \textbf{capped} & \textbf{uncapped} & \textbf{Software} & \textbf{capped} & \textbf{uncapped} \\
\midrule
Windows 10 & 4739 & 1026 & Onedrive & 815 & 0\\
Firefox & 2903 & 0 & Skype & 754 & 0\\
Office & 1943 & 277 & Zoom & 414 & 0\\
Dropbox & 1045 & 812 & Steam & 184 & 138\\
Chrome & 878 & 879  & Teamviewer & 87 & 0\\
Sophos & 861 & 861 & Nicehash & 1 & 1\\
Edge & 840 & 796 & Easyminer & 0 & 0\\
\bottomrule
\end{tabular}
}
\end{center}
\end{minipage}
\end{table}

We now briefly discuss the analysis results on the real-world RWTH dataset using rules extracted as described above, with and without a cap for \textit{minOpt}. The analysis took 3:55 hours, which equals to 15.300 packets analyzed per second. Note, that the dataset was recorded one year before rule generation, because we had no access to a more recent one. Hence, our rules were generated for newer versions of the applications. Also, because we extracted our rules from a small artificially generated dataset and test on a real-world dataset, our rules might be overfitting to the behavior of our VMs.

Table \ref{tab:realworld_results} shows DONUT's detected fingerprints in the real-world dataset. With capped \textit{minOpt}, we detected considerably more fingerprints. For some applications, the detections jumped from zero to several hundreds or thousands. We assume that our rules are overfitting because of the artificially generated data. In the capped scenario, we detected Windows on 11\% of all active hosts. Of the hosts where no application was detected, 48\% sent less than 100 DNS packets overall. Note, that the dataset contains desktops, laptops, and mobile phones, while we only created rules for Windows. We didn't find many fingerprints for crypto miners, which can be due to low popularity of these applications, or different behavior of the applications in our analyzed version compared to the real-world dataset. Our rule extraction only resulted in one label for Nicehash and three for Easyminer. The applications with more detections generally also queried more unique domains during data generation, resulting in more labels to work with. Nevertheless, DONUT still managed to detect a considerable amount of fingerprints in the dataset. Since we measured a low amount of false positives on our validation set, while changing the parameters mainly affected false negatives, we also assume a low false positive rate on this dataset.

\section{Conclusion and Future Work}
%In this work, we presented DONUT, a tool to fingerprint operating systems and applications including implications on the installed version based on \ac{DNS} traffic using a rule-based approach. DONUT and its rule syntax are designed to be modular and extensible for future additions.
In this paper, we presented the design and implementation of a pipeline for DNS-based application fingerprinting. We presented DONUT, a rule-based tool that analyzes DNS traffic in order to fingerprint software. Additionally, we developed ATLAS, a tool for automatically labeling network traffic based on the software that produced it, which we used to generate labeled datasets for 13 applications as well as Windows 10. Both tools will become open source alongside the publication of this paper. We automated the process of rule extraction based on data labeled by ATLAS, such that our pipeline can be easily extended to support more applications without much manual effort. Our performance evaluation shows that DONUT can analyze traffic of large networks at line-speed, independent of the number of rules or supported applications. We optimized the parameters used for automated rule-extraction and evaluated the functional correctness of DONUT with the help of an artificially generated dataset with known ground truth, resulting in low false positives and false negatives, even in edge-case scenarios. Additionally, we evaluated the results of DONUT on a large real-world dataset. 

\iffalse
\begin{itemize}
	\item Contributions:
	\item ATLAS labeling tool
	\item automated and scalable rule generation
	\item DONUT (without context-dependent and NAT detection functionality)
\end{itemize}
\fi

To evaluate our automated pipeline for application fingerprinting, the data from virtual machines was sufficient. For future work, we plan to collect more realistic and diverse data from real-world machines. As privacy concerns complicate the acquisition of labelled real-world data, we will simultaneously work on automating the interaction with applications to make the generation of artificial datasets more scalable. Also, we will further extend the functionality of ATLAS, by adding version information to the application labels in order to extract rules for different versions of applications. Finally, we plan to experiment with additional ways to parameterize the Set-rule extraction to further increase accuracy.

\iffalse
\begin{itemize}
	\item future work:
	\item automated application execution
	\item ATLAS support for other operating systems
	\item (include automated extraction of cd rules) <- weglassen?
	\item labeling of real world traffic with ATLAS, for rule extraction as well as evaluation
	\item automatically link multiple processes of one application together, e.g. by searching the installation folder
	\item evaluate on more recent dataset
\end{itemize}
\fi
%It is also desirable to improve DONUT's performance by adding more parallelization. For our implementation we use one thread for the entire analysis pipeline. It will not be possible to add inter-packet parallelization, because of inter-packet dependencies, but the \ac{CF}-Matcher and \ac{CD}-Matcher modules can still be changed to support multiple cores while processing one packet. 

%The generation of rules for fingerprinting additional operating systems and applications will increase DONUT's capabilities. Reliable version detection of programs also requires a deeper and more automated analysis. We manually generated the fingerprinting rules for DONUT, but for more additions, further automation the analysis and rule generation process is required. This includes automation of domain labeling, interval detection, sequence detection, and the generation of rules for DONUT. 

%We showed that DONUT can make a decisive contribution to network analysis and \ac{NAT} detection. It also shows the need of a more privacy preserving \ac{DNS} extension or replacement, due to the revealing nature of \ac{DNS}.
\appendix
\section{Ethics}
In this work, we used a real-world dataset containing DNS traffic captured on the edge routers of the RWTH university network to demonstrate the capabilities of our pipeline. Before we received access to the data, all IP addresses were anonymized with the tool \textit{capsan} to remove the possibility of linking domains to real users. In theory, the network administrator who was responsible for anonymizing the traffic can reverse this process. However, we received the dataset with a delay, such that mappings between IPs and users were already deleted once we had access to the anonymized data. Hence, even if the network administrator receives access to our analysis results, it is impossible to map any information to a specific user. Because the datasest might still contain sensitive information, it will not be made available to the public.
\section{Acknowledgments}
The authors would like to thank Jens Hektor for providing anonymized DNS data of the RWTH university network. This project has received funding from the European Union’s Horizon 2020 research and innovation programme under grant agreement No. 833418.

\bibliographystyle{ACM-Reference-Format}
\bibliography{bibliography}

\end{document}